\documentclass[10pt,amssymb, nofootinbib, superscriptaddress]{revtex4}
\usepackage{hyperref}
\usepackage{amsfonts,amssymb,amsmath,exscale,bbm}
\usepackage{footnpag} 
\usepackage[all,knot]{xy}  
\usepackage{color}
\usepackage{tikz} 
\usepackage{graphicx}

\newcommand\beq{\begin{equation}}
\newcommand\eeq{\end{equation}}
\newcommand\be{\begin{eqnarray}}
\newcommand\ee{\end{eqnarray}}
\newcommand\beqa{\begin{eqnarray}}
\newcommand\eeqa{\end{eqnarray}}
\newcommand\bean{\begin{eqnarray*}}
\newcommand\eean{\end{eqnarray*}}
\newcommand{\bes}{\begin{eqnarray}}
\newcommand{\ees}{\end{eqnarray}}

\newcommand{\cA}{{\mathcal A}}

\newcommand{\cG}{{\mathcal G}}

\newcommand{\cO}{{\mathcal O}}

\newcommand{\cS}{{\mathcal S}}

\newcommand\U{{\mathrm U}}

\newcommand\N{{\mathbb N}}
\newcommand\R{{\mathbb R}}
\newcommand\Z{{\mathbb Z}}

\newcommand{\SO}{{\rm SO}}
\newcommand{\so}{\mathfrak{so}}

\newcommand{\su}{\mathfrak{su}}
\newcommand{\SU}{{\rm SU}}

\def\inv{{\mbox{\tiny -1}}}
\def\minus{{\mbox{\small -}}}
\def\plus{{\mbox{\tiny +}}}

\newcounter{letter} \newcounter{numeral} \newcounter{Numeral}

\newcommand\Tr{\mathrm{Tr}}
\def\extd{\mathrm {d}}
\def\vphi{\varphi}
\def\vphihat{\widehat{\varphi}}
\newcommand\e{{\mbox{e}}}

\newcommand\acts\triangleright

\newcommand\maps{\colon}

\usepackage{hyperref}
 \usepackage{color}
\usepackage{amsfonts,amssymb,amsmath,exscale,bbm}
\usepackage{psfrag}
\usepackage{footnpag} 
\usepackage[all,knot]{xy}  
\usepackage{color}
\usepackage{tikz} 
\usepackage{graphicx}

\usepackage{amsmath}
\usepackage{amsfonts}
\usepackage{graphicx}
\usepackage{psfrag}
\usepackage{array}
\usepackage{bbm}
\usepackage{hyperref}
\usepackage{amsthm}
\theoremstyle{definition}

\def\inv{{\mbox{\tiny -1}}}

\def\minus{{\mbox{\small -}}}
\def\plus{{\mbox{\tiny +}}}

\def\hPsi{{\hat \Psi}}

\def\vphi{{\varphi}}

\def\dr{{\rightarrow}}

\def\hphi{{\widehat \phi}}

\def \hS{{\widehat{S}}}
\def \hC{{\widehat{C}}}
\def \hG{{\widehat{G}}}

\def\extd{\mathrm {d}}

\begin{document}

\title{\large \bf Quantum simplicial geometry in the group field theory formalism: \\ {reconsidering the Barrett-Crane model}}

\author{\bf Aristide Baratin\footnote{email: aristide.baratin@aei.mpg.de}}
\affiliation{Centre de Physique Th\'eorique,  CNRS UMR 7644, Ecole Polytechnique \\
 F-9112 Palaiseau Cedex, France}

\author{\bf Daniele Oriti\footnote{email: doriti@aei.mpg.de}} 
\affiliation{Max-Planck-Institut f\"ur Gravitationsphysik\\
Albert Einstein Institute\\
Am M\"uhlenberg 2, 14476, Golm, Germany, EU}



\begin{abstract}
A dual formulation of group field theories, obtained by a Fourier transform mapping functions on a group to functions on its Lie algebra, has been proposed recently. 
In the case of the Ooguri model for $\SO(4)$ BF theory, the variables of the dual field variables are thus $\so(4)$  bivectors, which have a direct interpretation as the discrete B variables.  
Here we study a modification of the model by means of a constraint operator implementing the simplicity of the bivectors, 
in such a way that projected fields describe metric tetrahedra. 
This involves a extension of the usual GFT framework,  where boundary operators are labelled by projected spin network states. 
By construction, the Feynman amplitudes are simplicial path integrals for constrained  BF theory. 
We show that the spin foam formulation of these amplitudes corresponds to a variant of the Barrett-Crane model for quantum gravity.  We then re-examin the arguments against the Barrett-Crane model(s), in light of our construction. 
\end{abstract}

\maketitle
\bigskip \bigskip

%
\section*{Introduction}
%

\noindent Group field theories (GFT) represent a second quantized framework for both spin networks and simplicial geometry \cite{iogft,iogft2, gftreview}, field theories on group manifolds (or Lie algebras) producing Feynman amplitudes which can be equivalently expressed as simplicial gravity path integrals \cite{aristidedaniele} or spin foam models \cite{mikecarlo}, in turn covariant formulations of spin networks dynamics \cite{LQG}\footnote{They can also be seen, from a more statistical field theory perspective, as a higher dimensional generalization of matrix models \cite{mm} and a particular class of tensor models \cite{tensor}}. Most spin foam (and GFT) models for 4d gravity are obtained starting from models describing topological BF theory and adding constraints on the 2-form B field, imposing it comes from a tetrad 1-form as $B= * e\wedge e$. This  is based on the Plebanski formulation \cite{constrainedBF,peldan} of General Relativity as a constrained BF theory. A classically equivalent modification of the same action is obtained by adding a topological term to it, which vanishes on-shell, with a coupling constant named Immirzi parameter \cite{peldan}. This gives the so-called Holst action, which is the classical starting point of loop quantum gravity (LQG)\cite{LQG}. 
The construction of spin foam amplitudes takes place in a simplicial context, by first assigning to the 2-dimensional faces of a simplicial complex data representing the discrete analogue of the fields of BF theory, that is Lie algebra elements representing the B field,  and group elements defining a discrete connection, and then defining a quantum amplitude imposing the Plebanski constraints on them. Ideally, one would want to constrain the discrete data at the classical level and then quantize the resulting geometric structures, as one would do in a simplicial path integral. However, this has proven very difficult up to now, and the usual route \cite{SF,BC,newSF} starts with a quantization of topological BF theory and proceeds by imposing the constraints at the quantum level, that is at the level of quantum states, hoping to obtain the correct result nevertheless, and despite the ambiguities involved in this way of proceeding. 
More precisely, the existing models \cite{newSF} (which exist in both euclidean and minkowskian signatures) result from the imposition of constraints via the solution of operator equations  (e.g. via a master constraint technique) or from constraining the quantum labels of BF coherent states, identified with the quantum analogue of the bivector fields of this theory, and are further distinguished by the presence and value of the Immirzi parameter. In absence of the Immirzi parameter $\gamma$ (on in the $\gamma=\infty$ sector of Holst theory), one gets the Barrett-Crane model \cite{BC} by using the operator method, and the so-called FK model \cite{newSF} via the coherent state method. With finite $\gamma>1$ the operator method gives the so-called EPRL model \cite{newSF} or an extension of the FK model. These two models remarkably coincide for $\gamma<1$, and in particular for $\gamma=0$ (corresponding to the \lq topological sector\rq of Holst gravity).  The new models thus succeed in incorporating the Immirzi parameter in the spin foam (and GFT \cite{BCgft1,BCgft2, vincentGFT-EPRL}) formalism, attempts at which had started very early  \cite{eteraImmirzi, danieleetera}.

\

In parallel with the above results, the simplicial path integral route has also been explored, providing important insights \cite{eteravalentin,thomasmuxin}. In particular, the analysis of \cite{eteravalentin} clarified to a great extent the (quantum) simplicial geometry of the Barrett-Crane model. The results we present in this paper can be seen as a new take on and a generalization of the results presented there. A decisive impulse to this line of research has been  given recently by the definition of a metric representation of GFTs \cite{aristidedaniele}\footnote{For earlier attempts aiming at the same result, see \cite{generalisedGFT}.}, in terms of B variables, and the proof of the exact duality of simplicial path integrals and spin foam models, within a GFT context, based on non-commutative tools \cite{PR3,laurentmajid,karim,etera}, then applied also to LQG states \cite{fluxes}. Despite being phrased in terms of simplicial geometry and in seemingly classical terms, as appropriate for a path integral quantization, also this method of quantization necessarily relies on a certain choice of quantization map (which dictates, for example, choices of operator orderings. The one that seems \cite{laurentmajid} to be at the root of the non-commutative Fourier transform and star product on which these results rely is the Duflo quantization map \cite{duflo}. This map is the most mathematically natural one for systems whose classical phase space is (based on) the cotangent bundle of some group manifold \cite{Alekseev}, has been successfully applied to the path integral formulation of Chern-Simons theory \cite{HannoThomas}, and argued to be relevant also for loop quantum gravity \cite{Alekseev}. Further support for the interpretation and use of these tools comes from their application to the much simpler case of the quantization of a particle on the sphere, where they allow to derive a complete and correct canonical path integral formulation of the dynamics \cite{Matti}\footnote{This metric formulation also allows to identify diffeomorphism transformations at the GFT level, and to link nicely their features across the canonical, spin foam and simplicial gravity formulation of the same theory \cite{GFTdiffeos}.}. 

\

\noindent 
The EPRL and FK models \cite{newSF} have now replaced the BC model in the interest of the spin foam community. On the one hand, this is due to the mentioned success in incorporating the Immirzi parameter in the formalism, and to the consequent close relation between the boundary states associated to these new models and those of canonical LQG. On the other hand, this is due to a variety of results and arguments that have been put forward as implying the failure of the Barrett-Crane model in correctly describing quantum geometry, and thus in representing a compelling candidate for a quantum gravity model\footnote{These arguments, mostly, refer to the BC {\it vertex} in the spin representation, and to the choice of data entering it, also because there is no definite consensus on the other ingredients  entering the model, i.e. \lq measure\rq     \, factors.}. 

\noindent These (numerous) arguments (related to each other) 
concern: the boundary states of the model (used also to define the vertex amplitude) and the quantization procedure used to obtain such states \cite{newSF}, the role of degenerate geometries \cite{asymBC2}, the possible lack of (discrete analogues of) needed secondary second class constraints \cite{sergei,sergei2}, and, most importantly, the encoding of simplicial geometry in the quantum amplitudes.

\

\noindent This last issue, often referred to as the \lq ultralocality of the BC model\rq  \quad can be articulated as:

a) 4-simplices speak only through face representations, i.e. triangle areas \cite{asymBC};

b) bivectors associated to same triangle in different simplices are not identified \cite{eteravalentin,newSF};


c) normal vectors to the same tetahedron seen in different 4-simplices are uncorrelated \cite{eteradanieleCoupling,aristidedaniele};

d) the simplicity constraints are imposed in a non-covariant fashion \cite{aristidedaniele};

e) because of this non-covariance, there are missing constraints over the connection variables \cite{aristidedaniele}.

\

The last two issues had been identified in \cite{aristidedaniele}, in a  first application of the non-commutative metric formalism to GFT gravity models, which has the advantage of making the simplicial geometry behind spin foam models manifest.
In this paper, we take further advantage of this representation to re-analyse the imposition of the simplicity constraints without Immirzi parameter in a GFT context, and thus in the simplicial gravity path integral based on BF theory appearing as Feynman amplitude of the theory. We actually {\it solve} the above two issues, and keep under control the rest of the encoding of simplicial geometry at the quantum level, by means of a simple generalization of the GFT formalism itself. This generalization has also the advantage of recasting the boundary states of the resulting model explicitly in the form of  {\it projected spin networks} \cite{projected, eterasergei}. Moreover, we show that the resulting model is {\it unique}, in a sense we will specify. The Feynman (spin foam) amplitudes turn out to be still a variant of the Barrett-Crane model. We are thus led to re-analyse the existing criticisms towards this model. 


While we leave the inclusion of the Immirzi parameter in the formalism
to a forthcoming paper \cite{new4dGFTImmirzi}, we also give a model for  the topological sector of Plebanski gravity and compare it to the so-called EPR spin foam model.  

We work in a GFT context because it is the most complete setting to study the dynamics of spin networks and simplicial structures. But it should be clear that the entire construction could be carried through directly at the level of simplicial path integrals (as done in \cite{eteravalentin}) or spin foam amplitudes, if one so prefers.

%
\section{GFT's in the metric representation} 

\noindent 
In this section, we briefly recall the construction of the metric formulation of GFT's \cite{aristidedaniele}. We start with the GFT model for 4d $\SO(4)$ BF theory, and then discuss, in the metric framework,  the implementation of the simplicity constraints \cite{aristidedaniele} leading to the (usual versions of the) Barrett-Crane model. 

\subsection{GFT for BF theory}
\label{ooguri}

The GFT for $\SO(4)$ BF theory is defined in terms of a field $\vphi_{1234}:= \vphi(g_1, \cdots g_4)$ on four copies of the group, satisfying the gauge invariance condition: 
\beq \label{gauge}
\forall  h \in \SO(4), \quad \vphi(g_1,\cdots  g_4) = \int \extd h \, \vphi(hg_1, hg_2, hg_3, hg_4)
\eeq
The dynamics is governed by the action:
\[
S =\! \frac{1}{2}\int [\extd g_i]^4 \varphi^2_{1234} -  \frac{\lambda}{5!} \! \int [\extd g_i]^{10}\varphi_{1234}\,\varphi_{4567}\,\varphi_{7389}\,\varphi_{962\,10}\,\varphi_{10\,851}.
\]
where $ [\extd g_i]^4$ and $[\extd g_i]^{10}$ denote the Haar measures on four and ten copies of $\SO(4)$. 
The combinatorics of the interaction term is that of a 4-simplex (the ten indices $i = 1, \cdots 10$ corresponding to its ten triangles),  while the kinetic terms dictates the gluing rules for 4-simplices along tetrahedra.
The quantum theory is defined by the perturbative expansion in $\lambda$ of the partition function and, by construction, the Feynman diagrams are 2-complexes dual to 4d simplicial complexes: vertices, links and faces (2-cells) of the graph are dual to 
4-simplices, tetrahedra and triangles of the simplicial complex, respectively. 
Using harmonic analysis on $\SO(4)$, the Feynman amplitudes take the form of the Ooguri state sum model \cite{Ooguri}. 


The metric formulation  \cite{aristidedaniele} of the same model is based on the group Fourier transform of the GFT field. 
The $\SO(3)$ group Fourier transform \cite{PR3,laurentmajid,karim, etera}, which maps functions on the group to functions on its Lie algebra $\su(2)$,  
straightforwardly extends to functions of (several copies of) $\SO(4)\sim \SU^-(2)\times \SU^+(2)/\mathbb{Z}_2$; it is invertible on even functions\footnote{In the sequel we thus assume the further invariance of the Ooguri field under $g_i \to -g_i$ in each of the variables.}
$f(g)\!=\!f(- g)$. The group Fourier transform of the field $\vphi$ is thus the function on four copies of the Lie algebra $\so(4)$ given by: 
\beq
\vphihat(x_1,..,x_4)\equiv \int[\extd g]^4\, \varphi(g_1,..,g_4)\, \e_{g_1}(x_1) \cdots \e_{g_4}(x_4),   \qquad x_i\in \so(4)\sim \R^6
\eeq
where $[\extd g]^4$ is the normalized Haar measure on $\SO(4)^4$. 
The plane waves $\e_g \maps \so(4) \to \U(1)$ are the functions defined as $\e_g(x) =  \e^{i \Tr x g}$, where $\Tr$ is trace is the fundamental representation of $\SO(4)$. Explicitly, using the decomposition of  the Lie algebra and group elements  $x=(x_-,x_+),  g = (g_-, g_+)$ into left  and right components 
$x_{\pm}\in \su(2), g_{\pm} \in \SU(2)$, and the parametrization $g_\pm \!=\! e^{\theta \vec{n}\cdot \vec{\tau}}$ and $x^\pm \!=\! \vec{x} \cdot \vec{\tau}$ of these components in terms of the (anti-Hermitian) $\su(2)$ generators $\vec{\tau} \!=(\! \tau_1, \tau_2, \tau_3)$, the plane waves read: 
\[
\e_g(x) = \e^{i\Tr x_\minus g_\minus}  \e^{i\Tr x_\plus g_\plus}
\]
where the trace is defined as $\Tr \tau_i \tau_j = - \delta_{ij}$. 

The space of dual fields inherits by duality a non-trivial (non-commutative) pointwise product from the convolution product on the group.  
It is defined on plane-waves as 
\beq  \label{star}
(\e_{g} \star \e_{g'})(x) \!:=\! \e_{gg'}(x),
\eeq 
extends component-wise to the tensor product of four plane-waves  and by linearity to the image of the Fourier transform.

The Lie algebra variables of the dual field $\vphihat$ are interpreted, geometrically, as the bivectors associated to a triangle, in each tetrahedron, in the standard discretized version of BF theory. 
Upon Fourier transform, the gauge invariance condition (\ref{gauge}) translates into a closure constraint $\widehat C$ for the bivectors associated to each tetrahedron (field), obtained as the group Fourier transform of the gauge invariance projector $\mathcal{P}$:
\beq \label{closure}
\mathcal{P} \varphi(g_1,..,g_4)= \int \extd h\, \varphi(hg_1,..,hg_4), \quad \dr \quad \widehat{ \mathcal{P} \varphi}= \widehat C\star\varphi,
 \eeq
 where $\widehat C(x_1,\cdots, x_4)= \delta_0(\sum_{n=1}^4 \,x_n)$, and  $\delta_0$ is the element $x=0$ of the family of functions $\delta_x$ defined as:
\beq \label{delta}
\delta_x(y)  := \int \extd h \, \e_{h^\inv}(x)  \e_h(y).
\eeq
These functions play the role of Dirac distributions in the non-commutative setting, as
\beq \label{Dirac}
\int \extd^6 y \, (\delta_x \star f)(y)= f(x) ,
\eeq
where $\extd^6 y$ is the standard Lebesgue measure on $\so(4) \sim \R^6$.

The combinatorial structure of the GFT action in the metric representation is the same as in group one, while group convolution is replaced by the $\star$-product. 
Using the short notation $\vphihat_{1234}:= \vphihat(x_1, \cdots, x_4)$, we can write the action as 
\beq \label{action in x}
S [\vphihat]  = \int [\extd^6 x_i]^4  \vphihat_{1234} \star \vphihat_{1234} 
+ \lambda  \int  [\extd^6 x_i]^{10}\, \vphihat_{1234} \star \vphihat_{4567}\star \vphihat_{7389} \star \vphihat_{96210}
\eeq
where it is understood that $\star$-products relate repeated lower indices as $\hphi_i \star \hphi_i \! :=\! (\hphi \star \hphi_{\minus})(x_i)$, with $\hphi_{\minus}(x) = \hphi(- x)$.  

In this representation, 
the Feynman amplitudes of the GFT can be computed by sticking together with the $\star$-product the
propagator and vertex functions given by:
\beq \label{Feynmanfunctions2}
P(x, x') = \int [\extd h]\, \prod_{i=1}^4 \delta_{\minus x_i}(x'_i),   \qquad 
V(x, x') = \int \prod_{\ell=1}^5 [\extd h_{\ell}]\, \prod_{i=1}^{10} ( \delta_{\minus x^{\ell}_i} \star\, \e_{h_\ell h^{\minus 1}_{\ell'}})(x^{\ell'}_i)
\eeq
The vertex function has twenty Lie algebra variables: two variables $x^\ell_i, x^{\ell'}_i$ for each  of the ten triangles of a 4-simplex, corresponding to the two tetrahedra $\ell, \ell'$ sharing that triangle. 
The vertex function encodes the identification of the bivectors associated to the same triangle in different tetrahedral frames \cite{aristidedaniele}, 
up to parallel transport between these frames given by the group elements $h_\ell h^{-1}_{\ell'}$;
the sign difference reflects the fact, in an oriented 4-simplex, a triangle inherit opposite orientations from the two tetrahedra sharing it. 

The Feynman amplitudes in this representation are simplicial path integrals for BF theory, involving an integration over one $\SO(4)$ variable $h_{l}$ for each link of the Feynman graph,  equivalently of each tetrahedron of the dual simplicial complex, which is interpreted as parallel transport between the two 4-simplices  sharing that tetrahedron; 
and  a single $\so(4)$ variable $x_t$  for each triangle $t$ of the dual simplicial complex:
\beq \label{bf} 
\cA_\cG = \int \prod_l \extd h_l\, \prod_e [\extd^6 x_t]  \, e^{i \sum_t \Tr \, x_t H_t},
\eeq
$\extd h_l$ is the Haar measure on $\SO(4)$ and  $\extd^6 x_t$ is the Lebesgue measure on $\so(4)\!\sim\!\R^6$. 
The group element $H_t:=\overrightarrow{\prod}_{l\in \partial f_t} h_l $ is the ordered product of group elements $h_l$ on the links on the boundary of the face (2-cell) $f_t$ of the graph dual to the triangle $t$, giving a measure of the discrete curvature associated to it. 
Such a discrete curvature is calculated for a given orientation of $f_t$, and a given 'reference' vertex in its boundary; the final amplitude is independent of this choice. 
The exponential corresponds to the product of plane waves $\prod_t \e_{H_t}(x_t)$; its argument $\sum_t \Tr x_t H_t$ is the unconstraint discrete BF action.

Importantly, the expression generalizes naturally \cite{aristidedaniele} to the case of open Feynman diagrams (GFT transition amplitudes), i.e. simplicial complexes with boundary, to give BF path integrals with fixed B variables on the boundary, with appropriate boundary terms \cite{danieleBoundary}.

\subsection{Imposing the simplicity constraints}
\label{BCusual}

This new representation of the Ooguri model for BF theory provides a convenient starting point for imposing in a geometrically transparent manner the discrete simplicity constraints that turn BF theory into 4d simplicial gravity. 
The simplicity conditions of Plebanski gravity \cite{DP-F} (or the gravitational sector of Holst gravity \cite{holst}), thus without any Immirzi parameter, 
ensure that the B-field can be expressed in terms of 1-forms fields $e_I$ as $B^{IJ} = \epsilon^{IJKL} e_I \wedge e_J$.
At the discrete level, part of these conditions (in their linearized form \cite{sergei,newSF,danielesteffen}) are implemented by requiring that, for each tetrahedron in the simplicial complex, 
the four bivectors $B^{IJ}_j$ associated to its four triangles $j=1, \cdots 4$, expressed in the reference frame of the tetrahedron, all lie in the hyperplane orthogonal to some normal vector in $\R^4$.
Concretely, using the canonical decomposition of bivectors into selfdual and anti-sefdual $\su(2)$-components $B^{\pm i} \tau_i$ with $2B^{\pm i} = \frac12 \epsilon^{ijk} B_{jk} + B^{0i}$, the conditions are imposed by requiring the existence of $k\in\!\SU(2) (\sim \cS^3)$ independent of $j$ 
such that:
\beq \label{linsimp}
\forall j, \quad B^+_j  = - k B^-_j k^{- 1}
\eeq

Back to GFT, where the role of the discrete $B$ are played by the field variables $x_j$ in the metric representation,  the condition (\ref{linsimp})  can be implemented on the GFT field by defining the projector $\vphihat \mapsto \hS_k \star \vphihat$, where  $\hS_k$ is the function of four $\so(4)$-variables defined in terms of the non-commutative $\delta$ functions (\ref{delta}) as 
\beq \label{simplicity}
\widehat{S}_k(x_1, \cdots x_4) =  \prod_{j=1}^4 \delta_{\minus kx^\minus_j k^\inv}(x^\plus_j)
\eeq
Such a projector imposes strongly the relations (\ref{linsimp})  on the field variables, since the non-commutative $\delta$ functions act as Dirac distribution for the $\star$-product.

Upon group Fourier transform, one can show that
\beq
\widehat{S}_k \star \vphihat= \int_{\SO(4)^4} [\extd g_j]^4\, (S_k\vphi)(g_jj) \,\e_{g_1}\cdots \e_{g_4}
\eeq
where $S_k$ projects onto fields on the product of four copies of the homogeneous space $\cS^3 \!\sim\! \SO(4)/ \SO(3)_k$, $\SO(3)_k$ being the stabilizer group of $k$ seen as a vector\footnote{Thanks to the isomorphism $\SO(4)\!=\!\SU(2)\!\times\! \SU(2) / \Z_2$, each element of $\SO(4)$ can be represented as $g\!=\!(g^\minus, g^\plus)$, with $g^\pm \in \SU(2)$.  
The map $\SO(4)\to \SU(2), \, g \mapsto g^\plus (g^\minus)^{\minus 1}$, with kernel $\SO(3) \!=\! \SU(2)/\Z_2$, realizes the identification of the homogeneous space 
$\cS^3 \sim \SO(4)/\SO(3)$ or set of normal vectors in $\R^4$, and the group $\SU(2)$. 
Through this identification, the rotation by $g\!=\!(g^\minus, g^\plus)$ of a normal vector represented by $k \in \SU(2)$ is represented by 
$g^{\plus} k (g^\minus)^{\minus 1}$. The stabilizer group $\SO(3)_k$ of $k$ is thus the set of $g\!=\!(g^\minus, g^\plus) \in \SO(4)$ such that 
$g^\minus = k^\inv g^\plus k$, ie of the form $g\!=\!(k^\inv uk, u)$, with $u \in \SU(2)$.}  of $\R^4$; namely, using the decomposition $g = (g^\minus, g^\plus)$ into self dual and anti-self dual components:
\[
(S_k \vphi)(g) := \int_{\SO(3)_k^4} [\extd u_j]^4 \, \vphi(k^\inv u_jk g_j^\minus, u_j g_j^\plus).
\]
The case $k\!=\!1$ reproduces the standard Barrett-Crane projector \cite{BCgft1,BCgft2}. This constraint is then imposed on each GFT field, i.e. on each tetrahedron of the simplicial complex, in its associated frame. The task of correctly parallel transporting these conditions in  the other frame associated to other tetrahedra and simplices of the simplicial complex is accomplished by propagator and vertex GFT functions.

The above amount to, roughly speaking, half of the full simplicity conditions ensuring geometricity of the bivectors. The other half, often linked to the secondary constraints arising from the canonical analysis \cite{sergei, biancajimmy}, can be shown to be imposed automatically if the closure condition is imposed on {\it all} tetrahedra in the simplicial complex (see, e.g. \cite{danielesteffen}), and if bivectors are correctly parallel transported across frames. This condition is imposed on GFT fields already at the level of the action, via the `closure' projector $\widehat{C}$ given by (\ref{closure}),  dual to the gauge invariance projector. 
Thus, it is imposed on {\it every} tetrahedron of the simplicial complex.  Therefore, we do not impose further conditions on the bivectors in what follows.

By combining the Barrett-Crane projector $\widehat{S}_1$ with closure, one can build up the field  
$\widehat{\Psi}\!:=\!\widehat{S}_1 \star \widehat{C}\star\vphihat$, defining a geometric tetrahedron. However, one immediately notices that, because the two functions $\widehat{S}_k$ and $\widehat{C}$ do not $\star$-commute, the product $\widehat{S}_k \star \widehat{C}$ does not act as a projector.  
This fact 
can be shown to be the manifestation of a geometric inconsistency in the way the resulting model imposes the simplicity constraints. 

In fact, given $h\!\in\!\SO(4)$, one has
\beq \label{commutation}
(\e_h \star \widehat{S}_k)(x) 
= (\widehat{S}_{h \rhd k} \star \e_h)(x)
\eeq
with $h \rhd k \!:=\!h^\plus k (h^\minus)^{\minus 1}$. This expresses the fact that, after rotation by $h$,  simple bivectors with respect to the normal $k$ become simple with respect to the {\sl rotated} normal $h \rhd k$. 
Imposing the simplicity condition everywhere with the same normal $k$ simply misses the fact that, under rotation $h$, normals should be rotated as well (which strengthens the relation between the normal variables and the gauge connection).
This means that the simplicity constraints are imposed {\it non-covariantly} by this procedure. 

But this is exactly this procedure that defines the usual versions of the Barrett-Crane models. In fact, combining the interaction term: 
\beq \label{BCvertex}
\frac{\lambda}{5!}\int \widehat{\Psi}_{1234}\star\widehat{\Psi}_{4567}\star\widehat{\Psi}_{7389}\star\widehat{\Psi}_{962\,10}\star\widehat{\Psi}_{10\,851}
\eeq
where $\widehat{\Psi}\!:=\!\widehat{S}_1 \star \widehat{C}\star\vphihat$, with the possible kinetic terms:
\beq \label{BCkinetic}
\frac{1}{2}\int \widehat{\Psi}^{\star2}_{1234}\; , \hspace{0.1cm} \frac{1}{2} \int (\widehat{C}\star\vphihat)^{\star2}_{1234}\, \hspace{0.1cm} \text{or}\hspace{0.2cm} \frac{1}{2} \int \vphihat^{\star2}_{1234}
\eeq
gives the versions of the Barrett-Crane model derived in \cite{BCgft1}, \cite{BCgft2} or \cite{eteravalentin} respectively. 
The existence of different versions, all differing, in spin foam representations, by edge (tetrahedra) amplitudes only,  is then understood to be due precisely to the non-covariant imposition of the simplicity constraints (see \cite{danieleBoundary}). 

In this metric representation, the GFT Feynman amplitudes of the constrained theories defined by (\ref{BCvertex}) and (\ref{BCkinetic}) take the form of simplicial gravity path integrals, with a measure weighted by a constrained BF action \cite{aristidedaniele}. A more general version of this calculation is detailed in the next section, so we do not repeat it here. The spin foam representation is then obtained by the successive application of the group Fourier transform and of the Peter-Weyl decomposition to the same amplitudes, and by the integration of all but the representation variables to encode the dynamics of the quantum simplicial geometry.




One can then analyze the encoding of simplicial geometric conditions in this model directly from the form of the GFT action, and test the criticisms mentioned in the introduction. Here we only notice that the model couples the bivector variables $x$ across simplices and that the interplay of simplicity constraints and parallel transport (following the covariance constraint) leads to further conditions involving both bivector and connection variables. It would be tempting to interpret them as a discrete version of the secondary constraints of the canonical theory, but this cannot be done in a rigorous way at the present stage. At the same time normals to the same tetrahedron, as seen in different 4-simplices, are {\it not} correlated, as they should on geometric grounds, and this in turn implies a missing geometric condition on connection variables $h_{\ell}$ -- the one resulting from the requirement that it correctly transports normal vectors across simplices.

We postpone a more detailed discussion of the simplicial geometric aspects to the last section. 
But first, we will introduce a modification of the above construction in order to solve the issue of non-covariant imposition of the simplicity constraints, and thus the consequent lack of constraints on the discrete connection and of correlations among normal variables. This is the goal of the next section.

\section{Extended GFT formalism}

\label{extended}

We have seen that some of the problematic issues in the standard formulation of the Barrett-Crane model have to do with the way normal vectors to each tetrahedron are treated. A simple generalization of the GFT field on which both the Ooguri and Barrett-Crane models are based  leads to an easy solution of them.

\subsection{Tetrahedra normals and gauge covariance}

The needed step is to promote the would-be normal vector to the tetrahedron to a dynamical variable $k\in S^3\simeq \SU(2)$, by adding it as an independent argument to the Ooguri field, representing the tetrahedron\footnote{Obviously, before any geometricity condition is imposed,  constraining the bivector variables to be triangle area bivectors and relating them to the normal vector, any geometric interpretation is to be understood as purely suggestive.}. Thus, in this extended formalism,  GFT are defined in terms of fields on 
$\so(4)^4 \times \SU(2)$: 
\beq
\vphihat_k(x_1, \cdots, x_4): = \vphihat(x_1, \cdots, x_4; k) = \int [\extd g]^4 \, \vphi(g_1, \cdots, g_4; k) \, \e_{g_1}(x_1) \e_{g_2}(x_2)\e_{g_3}(x_3)\e_{g_4}(x_4)
\eeq

In adopting the new field as our basic dynamical variable in a new formulation of quantum 4d BF theory, we still want to impose the local gauge invariance of the simplicial theory, as the invariance under change of local frame in each tetrahedron. This time, however, the local rotation should rotate simultaneously the bivectors and the normal vectors. In other words, the gauge invariance condition for the Ooguri field is now replaced by a gauge covariance of the $\SO(4)$ arguments with respect to the normal $k$, with the overall 5-argument field being invariant:
\beq \label{extgauge}
\vphi(h g_1, \cdots,h g_4; h \acts k) = \vphi(g_1, \cdots, g_4; k) \qquad \forall h \in  \SO(4)
\eeq
where $h \acts k:=h^{\plus} k (h^\minus)^{\minus 1}$ denotes the action of the $\SO(4)$-rotation $h\!:=\! (h^\minus, h^\plus)$ on the normal vector\footnote{See footnote on Sec. \ref{BCusual}.} $k\in \SU(2) \sim \cS^3$. 

To express this invariance in the metric representation, let us introduce the family  of functions labelled by $h \in \SO(4)$, 
given by a product of four $\SO(4)$ plane waves $\e_h(x):=\e_{h^\minus}(x^\minus) \e_h^\plus(x^\plus)$\footnote{Note that this is the same functions used to impose gauge invariance in the Ooguri BF model.}: 
\beq \label{prodpw}
\hC_h(x_1, \cdots x_4) = \e_h(x_1) \cdots \e_h(x_4)
\eeq
Upon group Fourier transform, the gauge covariance condition (\ref{extgauge}) reads
\beq
(\hC_h \star \vphihat_{ h^\inv \acts k})(x_1, \cdots, x_4) = \vphihat_k(x_1,\cdots,x_4)  \qquad \forall h \in  \SO(4)
\eeq
This condition is implemented by means of the constraint projector:
\beq  \label{extgaugemetric}
\hC \vphihat_k := \int \extd h \, \hC_h \star \vphihat_{h^\inv \acts k}
\eeq

Two important points must be noted, here. 
First, the $\SO(4)$-rotation invariance (\ref{extgauge}) of the extended field induces an invariance under the stabilizer group $\SO(3)_k$ of the normal $k$, 
affecting only the first four arguments of the GFT field. 
Upon Peter-Weyl decomposition of the $\vphi_k$, this means the fields modes are labelled by one normal vector $k$, 
four irreducible $\SO(4)$ representations (given by pairs of $\SU(2)$ spins $J_i = (j_i^\plus, j_i^\minus)$, $i=1\cdots 4$), each of which can be further decomposed into $\SO(3)_k$ representations $k_i$; and a four valent $\SO(3)_k$ intertwiner  contracting these representations. 
In terms of the dual spin network vertex, the labeling thus corresponds to a $\SO(4)$ representation in each link, and a normal vector $k$ and a $\SO(3)_k$ intertwiner
at the node. A set of basis functions is thus given by:
\beq \label{spinbasis}
\Psi^{(J_i, k_i, j)}_{m^\minus_i, m^\plus_i}(g_i;k) = \left(\prod_{i=1}^4 D^{j_i^\minus}_{n^\minus_i m^\minus_i}(g_i^\minus) 
D^{j_i^\plus}_{n_i^\plus m_i^\plus}(g_i^\plus) \widetilde{C}^{j_i^\minus j^\plus_i k_i}_{n^\minus_i n_i^\plus p_i}(k)\right)  (\iota_j)^{k_i}_{p_i}
\eeq
where repeated lower indices are  summed over. $D^{j^\pm}(g^\pm)$ are the $\SU(2)$ Wigner matrices, $(\iota_j)^{k_i}$ form a basis of four-valent $\SO(3)$ intertwiners, labelled by an intermediate spin $j$. 
The k-dependent coefficients, 
defined in terms of the 
$\SO(3)$ Clebsch-Gordan coefficients $C^{j^\minus j^\plus k}_{mnp}$ as:
\beq
\widetilde{C}^{j_i^\minus j^\plus_i k_i}_{m^\minus_i m_i^\plus p_i}(k) = \sum_{m} C^{j_i^\minus j^\plus_i k_i}_{m m_i^\plus p_i} D^{j^\minus_i}_{m m^\minus_i}(k)
\eeq
define a tensor that intertwines  the action of $\SO(3)_k$ in the representation $j^\minus_i \otimes j^\plus_i$ and the action of $\SO(3)$ in the representation $k_i$.
Namely, given $\mathbf{u}_k \!=\! (k^{\minus 1} u k, u)\!\in\! \SO(3)_k$, we have:
\beq
\widetilde{C}^{j_i^\minus j^\plus_i k_i}_{m^\minus_i m_i^\plus p_i}(k) D^{j^\minus_i}_{m^\minus_i n^\minus_i}(\mathbf{u}_k^\minus) D^{j^\plus_i}_{m^\plus n^\plus_i}(\mathbf{u}_k^\plus)
= \widetilde{C}^{j_i^\minus j^\plus_i k_i}_{n^\minus_i n_i^\plus q_i}(k) D^{k_{i}}_{q_i p_i}(u).
\eeq
(\ref{spinbasis}) corresponds to  the vertex  structure of the so-called projected spin networks of the covariant approach to loop quantum gravity \cite{projected}, 
which thus label the polynomial gauge invariant operators in the extended formalism. 
The convolution of GFT fields to define generic observables, and thus generic boundary spin network states at the level of transition amplitudes, will still be defined with respect to the $\SO(4)$ connection. The need for a generalization of the standard gauge invariance condition on quantum states (Gauss law) in the quantization of Plebanski gravity, and in the presence of linearized simplicity constraints, which then leads to projected spin network states, has been argued from a canonical continuum point of view in \cite{canPleb} (see also \cite{sergei, sergei2, danielesteffen}).

The second point is that, despite this generalization, it is easy to check that the following action of the invariant (under generalized gauge invariance) field $\vphi_k$:
\beq \label{extBF}
S[\vphi_k] =\! \frac{1}{2}\int \vphi^2_{1234,k} -  \frac{\lambda}{5!} \! \int \varphi_{1234,k_1}\,\varphi_{4567,k_2}\,\varphi_{7389,k_3}\,\varphi_{962\,10,k_4}\,\varphi_{10\,851k_5}.
\eeq
where we used the shorthand notation $\vphi_{1234,k}\!:=\!\vphi_k(g_1,\cdots, g_4)$, defines the same amplitudes as the original Ooguri model, thus still corresponds at the perturbative level to quantum simplicial BF theory.

\subsection{Simplicity constraints and geometricity projector}

To obtain a geometric theory in the extended framework,  we would like to impose the Plebanski simplicity constraints on the  model (\ref{extBF}).  
Just like in the previous section, one can implement these  constraints by means of the functions $\widehat{S}_k$ labeled by $k \in \SU(2)$, defined in (\ref{simplicity}),
where this time the label $k$ is coupled to the additional field variable. 
We thus define a simplicity projector $\vphihat \mapsto \hS \vphihat$  acting on the extended fields $\vphihat_k(x_1, \cdots x_4):=\vphihat(x_1, \cdots x_4; k)$ as: 
\beq
(\hS\vphihat)_k := \widehat{S}_k \star \vphihat_k
\eeq

Thus the only modification with respect to the GFT formulations of the Barrett-Crane model is that now the simplicity projections have to be defined imposing orthogonality of bivectors with respect to the normal vector that comes as a new argument of the field. This leads however to a very important difference with respect to the previous case.

In fact, remarkably, this simplicity projector commutes with the action of a rotation, in the sense that, 
given $h\!\in\!\SO(4)$,
\[
\hC_h \star (\widehat{S}\vphihat)_{h^\inv \acts k} = \hC_h \star \hS_{h^\inv \acts k} \star \vphihat_{h^\inv \acts k} =
\hS_k \star \hC_h \star \vphihat_{h^\inv\acts k} = \hS (\hC_h \star \vphihat_{h^\inv\acts k})
\]
where $h^{\minus 1}  \acts k:=h^{\minus} k (h^\plus)^{\minus 1}$ and $\hC_h$ are the functions of $\so(4)^4$ defined in (\ref{prodpw}). 
The first term correspond to the action of a rotation $h$ on the projected field $\widehat{S} \vphihat$; 
the last  term corresponds to the projection by $\hS$ of the rotated extended field $\hC_h \star \vphihat_{h^\inv\acts k}$. 
The simplicity projector acts on the rotated field by imposing simplicity with respect to the rotated normal vector. 
Thus, the equality just states that rotating by $h$ bivectors that are simple with respect to a normal $k$ gives bivectors that are simple with respect to the rotated normal
$h^{\minus 1}  \acts k$. 

This  commutation property is an important property, because it means that the model constructed in these terms will not suffer from the `lack of covariance'  noted in \cite{aristidedaniele} in the various versions of the Barrett-Crane model, and that the simplicity constraints are now going to be correctly imposed in the different frames across the simplicial complex, obtained as a Feynman diagram of the GFT model in the following section. 
  
This has a further crucial consequence. Simplicity and gauge covariance are encoded into two projectors $\hS$ and $\hC$, defined in (\ref{simplicity}) and (\ref{extgaugemetric}),  which are now commuting: 
their product therefore defines a {\it projector} $\hG\!=\! \hS\hC \!=\! \hC\hS$, with $G^2=1$. 
Explicitly, in the metric representation, this projector acts as:  
\beq
(\hG  \vphihat)(x_j; k) = \int_{\SO(4)} \extd h \left[\prod_{j=1}^4 \delta_{\minus kx^\minus_j k^\inv}(x^\plus_j)\right] \star \left[\e_h(x_1)\cdots \e_h(x_4)\right]  \star 
\vphi(x_1, \cdots x_4; h^{\minus} k (h^\plus)^{\minus 1})
\eeq
Upon group Fourier transform, it acts dually on extended group fields as
\beq
(G \vphi)(g_j; k) = \int_{\SO(4)} \! \extd h \left(\prod_j  \int_{\SU(2)} \extd u_j\right)
\vphi((h^\minus k^\inv u_j k g_j^\minus, h^\plus u_j g_j^\plus); h^\plus k (h^\minus)^{\minus 1})
\eeq
The geometric nature of the tetrahedron corresponding to the GFT field $\vphihat$ is then enforced by the single projector $G$ that we can call {\it geometricity projector}. It is then imposing this projector on the generalized fields $\vphihat$ that one can construct a GFT (and spin foam) model of 4d quantum gravity based on the (linear) Plebanski formulation of classical gravity, free of the mentioned problematic features (from the point of view of simplicial geometry) of the usual formulations of the Barrett-Crane model. This is what we do in the following section.

\section{A revised model for 4d quantum gravity with no Immirzi parameter}

\subsection{The GFT model and its Feynman amplitudes}

We start from the extended Ooguri model introduced in section \ref{extended}, and use the geometricity projector $G$ combining both (extended) closure and simplicity constraints in a single operator. We define a projected field  $\widehat{\Psi}:= \hG  \vphihat$, representing a geometric (non-commutative) tetrahedron, whose geometry is characterized by four area bivectors $x_j$ and one normal vector $k$. 

The GFT model is defined by the action:
 
\beq \label{action}
S(\widehat{\Psi})\,=\,\frac{1}{2}\int [\extd^6 x_i]^4 \int \extd k \, \widehat{\Psi}^{\star2}_{1234, k}\,+\,\frac{\lambda}{5!}\int [\extd^{6} x_i]^{10} [\extd k_a]^5\, \widehat{\Psi}_{1234, k_a}\star\widehat{\Psi}_{4567, k_b}\star\widehat{\Psi}_{7389, k_c}\star\widehat{\Psi}_{962\,10, k_d}\star\widehat{\Psi}_{10\,851, k_e}\, ,
\eeq
where $\extd^6 x_i$ is the Lebesgue measure on $\so(4) \sim \R^6$ and $\extd k$ is the Haar measure on $\SU(2)$. It is understood, just like in the previous sections,  that $\star$-products relate repeated lower  indices as $\hPsi_i \star \hPsi_i \! :=\! (\hPsi \star \hPsi_{\minus})(x_i)$, with $\hPsi_{\minus}(x) = \hPsi(- x)$.   

We can immediately notice four main features of the model so defined:
\begin{itemize}
\item Just as in the Ooguri model and in the standard Barrett-Crane model(s), bivector variables associated to the same triangle in different tetrahedra (fields) within the same 4-simplex, as well as across different 4-simplices, are identified in the frame of the same 4-simplex, to which the above writing refers. When expanding the gauge covariance constraint in integral form and writing explicitly the delta functions relating the arguments of the fields, one sees that the same bivector variables are now expressed in the frames associated to the various tetrahedra, and are related by the discrete connection introduced by covariance constraint (see also \cite{aristidedaniele,generalisedGFT,biancajimmy,danielesteffen}).

\item Just as in the extended Ooguri model, normal vectors to tetrahedra are coupled only indirectly in each 4-simplex, through their relation with bivector variables imposed by the generalized covariance condition and by the simplicity constraints $\widehat{S}_k$. On the other hand, the normal vectors associated to the same tetrahedron in the two different 4-simplices sharing it are strictly identified by the kinetic term {\it in the reference frame associated to the tetrahedron} (this becomes apparent when making explicit the form of the interaction and kinetic functions, and writing the generalized closure constraint in integral form).

\item The use of the generalized geometricity projector $\hG$ imposes all the necessary geometric conditions on the bivector variables in each tetrahedron. 
Note that the closure of the four field variables, hence of the bivectors associated to the four triangles of a tetrahedron, is recovered by integration over the normals. In fact, such integration makes the insertion of a closure constraint  in front of each field $\hPsi_{1234,k}$ in the interaction term redundant, since:
\beq
\int \extd k \delta(x_1 + \cdots + x_4) \star \hPsi(x_1,\!\cdots\!, x_4; k) = \int \extd k  \hPsi(x_1, \!\cdots\!, x_4;k)
\eeq

\item The nature of the operator $\hG$ as the product of two commuting projectors leads to a unique definition of the constrained model. 
By contrast, the standard Barrett-Crane projector does not commute with gauge invariance, giving rise to some ambiguities in the form of 
edge amplitudes of the spin foam model. 
Here,  for example, the use of the interaction term of (\ref{action}) with any of the kinetic terms:
\[
\frac{1}{2}\int \widehat{\Psi}^{\star2}_{1234, k}\; , \hspace{0.1cm} \frac{1}{2} \int (C\acts \vphihat)^{\star2}_{1234, k}\, ,\frac{1}{2} \int (\hS_k\acts \vphihat)^{\star2}_{1234, k} \hspace{0.1cm} \text{or}\hspace{0.2cm} \frac{1}{2} \int \vphihat^{\star2}_{1234, k}
\]
leads to the same amplitudes upon perturbative expansion. At the same time it should be stressed that any of the above choices would lead, strictly speaking, to the definition of a different field theory, for what concerns the related space of fields, the classical equations of motion, etc.
\end{itemize}

\

The Feynman amplitudes of the model (\ref{action}), in the metric representation, can be computed with the propagator and vertex functions given by\footnote{This corresponds to the choice: $S(\vphihat)\,=\,\frac{1}{2}\int \vphihat^{\star2}_{1234, k}\,+\,\frac{\lambda}{5!}\int  \widehat{\Psi}_{1234, k_a}\star\widehat{\Psi}_{4567, k_b}\star\widehat{\Psi}_{7389, k_c}\star\widehat{\Psi}_{962\,10, k_d}\star\widehat{\Psi}_{10\,851, k_e}$}:
\beq \label{FeynmanfunctionsBC}
P(x_i, x'_i; k, k') =  \delta_{\minus x_i}(x'_i) \, \delta(k' k^{\minus 1}) \qquad
V(x^\ell_i, x^{\ell'}_i; k_\ell, k_{\ell'}) = \int \prod_{\ell=1}^4 \extd h_{\ell} \prod_{i=1}^6( \delta_{\minus x^\ell_i} \star S_{k_{\ell}} \star \e_{h_{\ell} h_{\ell'}^\inv} \star S_{k_{\ell'}})(x_i^{\ell'})
\eeq
where $S_k(x) := \delta_{\minus kx^\minus k^\inv}(x^\plus)$ and $\extd h_{\ell}$ is the Haar measure on $\SO(4)$. 
Note that the vertex does not encode any correlation among normal vectors for different tetrahedra. The propagator, instead, presents a fifth strand corresponding to the identification of normal vectors in the two 4-simplices sharing the same tetrahedron. 


The calculation of the Feynman amplitudes  is similar to the topological case \cite{aristidedaniele}. 
In building up the diagram, propagator and vertex strands are joined to one another using the $\star$-product. 
The amplitudes are integrals over the bivector and holonomy variables, and the normals.  
In terms of the simplicial complex dual to the Feynman graph, bivectors $x_t^\tau$ are labelled by a couple $\{$triangle $t$, tetrahedron $\tau\}$ with $t\subset \tau$;
holonomies $h_{\sigma\tau}$ are labelled as by a couple $\{$4-simplex $\sigma$, tetrahedron $\tau\}$ with 
$\tau\subset \sigma$; the normals are labelled by tetrahedra as $k_\tau$. 

Under the integration over the normals and holonomy variables,  the amplitude of a closed graph factorizes into a product of contributions $\cA_{t}[h_{\sigma\tau}, k_{\tau}]$ for each {\sl face} of the graph, hence for each {\sl triangle} $t$ of the dual simplicial complex, which take the form a cyclic $\star$-product:  
\beq 
\cA_{t}[h_{\sigma\tau}, k_{\tau}] = \int \prod_{j=0}^{N_t} [\extd^6 x^{j}_t]\, \vec{\bigstar}_{j=0}^{N_t} \, (\delta_{x^{j}_t} \star S_{k_{j}} \star \e_{h_{jj\plus1}})(x^{j+1}_t) 
\eeq
The integer $j \! \in\! \{0\cdots N_t\}$ labels the tetrahedra sharing the triangle $t$, or the links of the boundary of the face $f_t$ dual to $t$ in the Feynman graph. The expression is defined for a given ordering of these tetrahedra, induced by a choice of orientation of $f_t$ and a reference point in its boundary. 
$h_{jj\plus1} := h^{\minus 1}_{\sigma j} h_{\sigma j\plus1 }$ is the holonomy from the tetrahedron $j$ to the tetrahedron $j+1$ through  the 4-simplex $\sigma$ sharing them.  By convention we set $x_{N+1} :=x_0$. 
The variables $x^j_t \in \so(4)$ corresponds to the same bivector associated to the triangle $t$, expressed  in the various reference frames of the tetrahedra $j$. 
The measure features both the simplicity constraints for the  bivector at the level of each tetrahedron and the parallel transports of (constrained) bivectors and normal vectors across simplices, via the discrete connection. 


The Feynman amplitude can be written as a simplicial path integral, by first integrating, in each face contribution $\cA_{t}[h_{\sigma\tau}, k_{\tau}]$,  over all bivectors $x_t^j$ save one 
$x_t\!:=\!x_0$ corresponding to a `reference' tetrahedron frame $j=0$; 
and by a repeat use of the commutation relation $\e_h \star S_k = S_{h \acts k}  \star \e_h $, with  $h \acts k = h^\plus k h^{\minus 1}$,  
to place all the plane waves to the right of the integrand. 
We obtain the following expression for the graph amplitude: 
\beq \label{simplicial-graphamp}
\cA_\cG = \int \prod_{<\sigma\tau>} \extd h_{\sigma\tau} \prod_t \extd^6x_t  \left[\int \prod_{\tau} \extd k_{\tau} \, 
\prod_t \vec{\bigstar}_{j=0}^{N_t}  \, S_{h_{0j} \acts k_j}(x_t)  \right] \star e^{i \sum_t \Tr \, x_t H_t}
\eeq
where  the holonomy $h_{0j} = h_{01}  \cdots h_{j\minus 1 j}$ parallel transports the reference tetrahedron 0 to the tetrahedron $j$ (with $h_{00} = 1$) and 
$H_t := h_{01} \cdots h_{N0}$ is the holonomy along the boundary of the face $f_t$ of $\cG$ dual to the triangle $t$, calculated from the reference tetrahedron frame. 
The  measures $\extd h_{\tau\sigma}, \extd^6 x_i$ and $\extd k_\tau$ are respectively the Haar measure on $\SO(4)$, the Lebesgue measure on $\so(4) \sim \R^6$ and the Haar measure on $\SU(2)$. The exponential corresponds to the product of plane waves $\prod_t \e_{H_t}(x_t)$; its argument $\sum_t \Tr x_t H_t$ is the unconstraint discrete BF action.

This amplitude takes the form of a (non-commutative) simplicial path integral for a constrained BF theory. The measure features the simplicity of the bivector $x_t$ of a triangle in the frame of all tetrahedra sharing the triangle. Thus, $S_{h_{0j} \acts k_j}(x_t)$ imposes the simplicity of $x_t$ with respect to the $h_{0j} \acts k_j$, which is the normal to the tetrahedron $j$ parallel-transported to the reference frame $0$. This is the same as requiring the simplicity with respect to $k_j$ of the 
bivector parallel-transported to the frame of $j$.
 
Note that the above amplitude automatically encodes the closure constraint for the bivectors within each tetrahedron.
This is a direct consequence of (extended) gauge invariance and the integration over the normals. Let us nevertheless check it explicitly, by
picking a tetrahedron $\tau$ and considering the variables $x_{t_i}, i=1\cdots 4$ associated to its four boundary triangles, namely,  bivectors expressed the reference tetrahedron frame of each triangle. The closure constraint says the sum of the four bivectors, once parallel-transported to the frame of $\tau$, is zero; in the non-commutative formalism, it is imposed using the non-commutative delta-function (\ref{delta}).  
Using suitable  gauge transformation, we can always suppose that $\tau$ is the reference tetrahedron for triangles $t_i$.  
By appropriately flipping signs $x_{t_i}\mapsto - x_{t_i}$, we can also assume that the orientations of the faces $f_{t_i}$ chosen to calculate the holonomies induce the same orientation of the common link dual to $\tau$. Let us insert the closure constraint $\delta(\sum_{i} x_{t_i})\!:=\! \int \extd g  \prod_{i} \e_g(x_i)$ on the left of the simplicity functions in the integrand of (\ref{simplicial-graphamp}), using the star product. 
The term depending on $x_{t_i}$ in this integrand  then becomes:
\[
\int\extd g   \prod_{i=1}^4 \e_g(x_i) \star  \left[ \vec{\bigstar}_{j=0}^{N_{t_i}}  \, S_{h_{0j} \acts k_j}(x_{t_i})\right]  \star e^{i \sum_i \Tr \, x_{t_i} H_{t_i}}
=\int \extd g \prod_{i=1}^4  \vec{\bigstar}_{j=0}^{N_{t_i}}  S_{gh_{0j} \acts k_j}(x_{t_i})  \star e^{i \sum_i \Tr \, x_{t_i} g H_{t_i}}
\]
where the equality follows from the commutation relation (\ref{commutation}) between plane waves and simplicity functions and the definition of star product of plane waves. Now, the dependence upon $g$ in the integrand can be removed by the  change of variables
$k_\tau \mapsto g k_\tau, \, h_{\sigma_\tau \tau}\mapsto h_{\sigma_\tau \tau} g^{\minus 1} $,  where 
$\sigma_\tau$ is the 4-simplex\footnote{$\sigma_\tau$ is also the 4-simplex the dual to the `target' vertex of the oriented link dual to $\tau$.} to which both $\tau$  and its successor in the ordered lists $j=0 \cdots N_{t_i}$ of tetrahedra sharing $t_i$, belong.  
This shows that the insertion of the closure constraint is redundant, hence that such constraints are already implemented in the simplicial path integral.

The integrand of (\ref{simplicial-graphamp}),  as a function of the bivectors, holonomies and normals, is invariant under the following gauge transformations, generated by an $\SO(4)$ element $g_\tau$ for each tetrahedron:
\beq
h_{\sigma \tau} \mapsto h_{\sigma \tau} g^{\minus 1}_\tau,\qquad
k_\tau \mapsto g^\plus_\tau k_{\tau} (g_{\tau}^{\minus})^{\minus 1}\qquad
x_t \mapsto g^{\minus 1}_{\tau(t)} x_t g_{\tau(t)}
\eeq
where $\tau(t)$ is the reference tetrahedron for the triangle $t$ and $g_\tau^{\pm}$ arises from the selfdual/antiselfdual decomposition of $g_{\tau}$. 
This corresponds $\SO(4)$ rotation of  the tetrahedral frames. 
The integrand is also invariant under rotation of the simplex frames acting only on holonomies  $h_{\sigma \tau} \mapsto g_{\sigma} h_{\sigma \tau}$ (all the other variables being expressed in tetrahedral frames). 
Choosing $g_{\tau} \!=\! (g^\plus_\tau, g^\minus_{\tau}) \!:=\! (1, k^{\minus 1}_\tau)$ as gauge parameters corresponds to the `time gauge', which  fixes the normals to the north pole of $S^3$.  This shows that the integrals over the normals $k_{\tau}$ drop out of the amplitude (\ref{simplicial-graphamp}). 
Making explicit the simplicity functions and denoting $\bar{h}_{0j}:= h_{0j}^\plus (h_{0j}^\minus)^{\minus 1}$, we obtain: 
\beq \label{finalformamp}
\cA_\cG  = \int \prod_{<\sigma \tau >} \extd h_{\sigma \tau } \prod_t \extd^6x_t \left[ \prod_t \vec{\bigstar}_{j=0}^{N_t}  \, \delta_{
\minus \bar{h}_{0j} x_t^\minus \bar{h}_{0j}^\inv}(x_t^\plus) \right] \star e^{i \sum_t \Tr \, x_t H_t}
\eeq
Note that gauge fixing the normals is only possible for amplitudes of {\sl closed} graphs; the amplitude of open graphs, dual to simplicial complexes with boundary, has an explicit dependence on the normals to the boundary tetrahedra. We illustrate this in the case of the simplex boundary state in Sec. \ref{sec:boundarystate}. 
 
\

To summarize, we have written  the GFT  Feynman amplitudes as simplicial path integrals with the explicit form of a constrained BF theory, with a clear form of (linear) simplicity constraints imposed in each tetrahedral frame, with respect to a given normal variable and correctly parallel transported across frames by means of the gauge connection. The integration over the same connection imposes the closure constraint in every tetrahedron; this in turn implies the remaining (volume) simplicity constraints \cite{newSF,danielesteffen}. The amplitude also encodes constraints on the connection $\{h_{\sigma \tau}\}$,  following from simplicity, giving  a specific measure on the space of connections.



 \subsection{Spin foam representation of the amplitudes}
 \label{spinfoam}

The same amplitudes can be written in spin foam representation, exploiting the exact duality between the metric and spin representation of the GFT \cite{aristidedaniele}. This re-writing is obtained by inverse group Fourier transform -- giving the pure lattice gauge theory formulation of the model involving only group elements \cite{SF})  
-- and  the Peter-Weyl decomposition of the amplitudes,  leaving only a sum over representation of pure spin foam amplitudes.

For the present model the calculation goes as follows. 
We first notice that, under the integration of the group elements $h_{\sigma \tau}$ in  (\ref{finalformamp}), each integral over $x_t$ takes the form of an inverse SO(4) Fourier transform formula. Indeed, for a given set of holonomy variables  $\{h_{\sigma \tau}\}$, 
the functions $\vec{\bigstar}_{j=0}^{N_t}  \, \delta_{\minus \bar{h}_{0j} x_t^\minus \bar{h}_{0j}^\inv}(x_t^\plus)$
inside the brackets, labelled by $t$, define functions of $\su(2)$,  image by non-commutative Fourier transform of some functions $\cO_t$ on $\SO(4)$:
\[
 \vec{\bigstar}_{j=0}^{N_t}  \, \delta_{\minus \bar{h}_{0j} x_t^\minus \bar{h}_{0j}^\inv}(x_t^\plus) = \int_{\SO(4)} \extd g \cO_t(g) \e_g(x)
\]
The inverse Fourier transform formula (see for e.g \cite{aristidedaniele}) gives: 
\beq
\cO_t(g) = \int \extd^3 x \vec{\bigstar}_{j=0}^{N_t}  \, \delta_{\minus \bar{h}_{0j} x_t^\minus \bar{h}_{0j}^\inv}(x_t^\plus)  \star \e_{g^\inv}(x)
\eeq
A quick comparison between this formula and  the integrand of (\ref{finalformamp}) leads to the following simple expression of the graph amplitude  in terms of the group functions $\cO_t$:
\beq \label{groupamp}
\cA_\cG = \int \prod_{\tau \sigma} \extd h_{\tau \sigma} \prod_t \cO_t(H_t^{- 1})
\eeq
where we recall that $H_t:= h_{01} \cdots h_{N0}$ is the holonomy around the triangle $t$. 

Let us determine the functions $\cO_t$. The  plane-wave expansion of the non-commutative delta-functions gives: 
\beqa
 \vec{\bigstar}_{j=0}^{N}  \, \delta_{\minus \bar{h}_{0j} x_t^\minus \bar{h}_{0j}^\inv}(x_t^\plus) &=& \int \extd g \int \prod_{j=1}^{N} \extd g_j \, 
\e_{g \prod_{j=1}^{N} h_{0j}^\minus g_j h_{0j}^{\minus \inv}}(x_\minus) \, \e_{g \prod_{j=1}^{N} h_{0j}^\plus g_j h_{0j}^{\plus \inv} }(x_\plus)  \nonumber \\
\label{FourierOt}
&=& \int \prod_{\pm} \extd g_\pm \!\int \prod_{j=1}^{N} \extd g_j\,  \delta(g_\plus g_\minus^{\minus 1}) 
\e_{g_\minus \prod_{j=1}^{N} h_{0j}^\minus g_j h_{0j}^{\minus \inv}}(x_\minus) \, \e_{g_\plus \prod_{j=1}^{N} h_{0j}^\plus g_j h_{0j}^{\plus \inv} }(x_\plus)
\eeqa
where the integrations are over $\SU(2)$. We omitted the $t$ of $N_t$ to simplify the notations. After the following change of variables:
\beq
g_{\minus} \,\, \mapsto \,\, g_\minus \prod_{j=1}^N h_{0j}^{\minus } g_j h_{0j}^{\minus \inv}, 
\qquad g_\plus \,\, \mapsto  \,\, g_\plus \prod_{j=1}^N h_{0j}^{\plus }  g_jh_{0j}^{\plus \inv}
\eeq
the expression (\ref{FourierOt})  takes the explicit form of a $\SO(4)$ Fourier transform
$\widehat{\cO}_t(x) \!=\! \int \extd g \cO_t(g) \e_{g}(x)$, where $\cO_t(g)$ is given by: 
\beq
\cO_t(g) = \int \prod_{j=1}^N \extd g_j\, 
\delta(g_\plus\, \left[\prod_{j=1}^N h_{0j}^{\plus }  g_jh_{0j}^{\plus \inv}\right]^{\minus 1} \prod_{j=1}^N h_{0j}^{\minus } g_j h_{0j}^{\minus \inv} \, g_{\minus}^{\minus1} ) 
\eeq
The evaluation of this function for  $g = H_t^{-1}$ gives:
\beq \label{eval}
\cO_t(H_t^{-1}) = \int \prod_{j=1}^N \extd g_j\, 
\delta(\left[h_{01}^{ \plus} g_1 h_{12}^{\plus} \cdots g_N h_{N0}^{\plus}\right]^{\minus1} h_{01}^\minus g_1 h_{12}^\minus \cdots g_N h_{N0}^\minus)
\eeq
Next, we use the Plancherel decomposition of the $\SU(2)$ delta function in terms of the characters $\chi^J(g) \!=\! \Tr D^J(g)$ in the $\SU(2)$-representations labelled by $J \in \frac12 \N$, which also labels the simple $\SO(4)$ representation $(J,J)$ \cite{BC,mi}, to write: 
\beq
\cO_t(H_t^{-1}) = \sum_{J } d_J \int \prod_{j=1}^N \extd g_j\, \chi^J(\left[h_{01}^{ \plus} g_1 h_{12}^{\plus} \cdots g_N h_{N0}^{\plus}\right]^{\minus1} h_{01}^\minus g_1 h_{12}^\minus \cdots g_N h_{N0}^\minus)
\eeq
where $d_J \!=\! 2J\!+\!1$. 
The integration over each group variables $g_j$ in (\ref{eval}), which appears exactly two times in the argument of each character $\chi^J$, is performed using $N$ times the orthogonality relation of the representation matrices \[\int \extd g D^J_{mn}(g^{\minus 1}) D^J_{pq}(g) = \frac{1}{d_J} \delta_{m, q} \delta_{n, p}\] 
In fact, if we denote $\chi^J_0:= \chi^J(h_{01}^{\plus} (h_{01}^{\minus})^{\minus1}$ and, for $K=1 \cdots N$:
\[
\chi^J_K:=\chi^J(\left[h_{01}^{ \plus} g_1 h_{12}^{\plus} \cdots g_K h_{KK\plus1}^{\plus}\right]^{\minus1} h_{01}^\minus g_1 h_{12}^\minus \cdots g_K h_{KK\plus1}^\minus),
\]
(where by convention $N+1=0$), one can show, using orthogonality, the following recursion relations:  
\beq
\int \extd g_K  \chi^J_K = \frac{1}{d_J} \chi^J(h_{KK\plus1}^{\plus}( h_{KK\plus1}^{\minus})^{\minus 1}) \chi^J_{K-1}
\eeq
We conclude by iteration: 
\beq
\cO_t(H_t^{-1}) = \sum_{J } \frac{d^2_J}{d_J^{N+1}}\prod_{j=0}^N \chi^J(h_{jj\plus1}^{\plus} (h_{jj\plus1}^{\minus})^{\minus 1})
\eeq
where $N\!+\!1$  is the number of tetrahedra sharing the triangle $t$. 

Using the form (\ref{groupamp}) of  the Feynman amplitude (\ref{finalformamp}), and the above expression for the functions $\cO_t$,
we finally obtain the spin foam amplitude: 
\beq
\cA_\cG = \sum_{\{J_t\}} \prod_t d^2_{J_t} \prod_\tau \frac{1}{\prod_{t \in \partial\tau} d_{J_t}}  \prod_{\sigma} \{10J\}_{\sigma} 
\eeq
The sum is over the $\SU(2)$ spins $J_t$ labeling the triangles of the simplicial complex dual to $\cG$  and the products are over the triangles $t$, the tetrahedra $\tau$ and the 4-simplices $\sigma$. The 4-simplex weight  $\{10J\}_{\sigma}$  is the Barrett-Crane $10j$-symbol \cite{mi}:
\beq
\{10J\}_{\sigma}  = \int \prod_{v} \extd h_{v} \prod_{l=(vv')} \chi^{J_{t(l)}}(h^\plus_{v} (h_{v'}^\minus)^{\minus 1})
\eeq
In this expression, $v$ label the five vertices and $l$ the ten links of the graph dual to the boundary of the 4-simplex: 
a vertex is dual to a boundary tetrahedron, a link $l=(vv')$ is dual to a triangle $t(l)$ sharing two tetrahedra. The measure is over $\SO(4)$ elements $h_v\!:=\!(h_v^\plus, h_v^\minus)$. 

The model is thus a variant of the Barrett-Crane model  with a specific edge amplitude, different from that of other versions of the Barrett-Crane model \cite{danieleBoundary} (notice in particular the absence of the norm of the BC intertwiner from the amplitudes), although identical to that obtained in \cite{eteravalentin}, whose construction is indeed close to ours regarding both its motivations and techniques employed.  

From both the construction and the detail of the calculation, it should be clear that, while the appearance of the $10j$-symbol is the direct result of the form of the simplicity constraints chosen (corresponding to the pure gravity sector of the Holst action, or to the Plebanski formulation of gravity, with linear constraints), most of the information concerning the details of the simplicial geometry behind the model, in particular the extended closure condition and associated Lorentz covariance of quantum states, as well as the correlations of normal vectors across simplices, is now hidden in the edge amplitudes. The reason for this non-transparent form is obviously the fact that all the geometric variables in which this information is manifest, i.e. bivectors, gauge connection and normal vectors, have been integrated out.

\subsection{Simplex boundary state} \label{sec:boundarystate}

Taking the GFT Feynman amplitudes as the definition of a lattice model, we would like to discuss here the simplex boundary state. 
The vertex function (\ref{FeynmanfunctionsBC}) gives the amplitude for a single 4-simplex (represented by the vertex graph) 
with fixed boundary data $\{x_t^\tau, x_t^{\tau'}\}$ and $\{k_\tau\}$: the pair $x_t^\tau, x_t^{\tau'}$ both corresponds to
bivectors associated  the triangle $t$, expressed in the frame of the two tetrahedra sharing it. 
Given ten of such pairs, the vertex function encodes the relations between the bivectors of the same triangle expressed in different tetrahedral frames:
\beq \label{vertex}
\int \prod_\tau \extd h_{\tau} \prod_{t=<\tau \tau'>} ( \delta_{\minus x^{\tau}_t} \star S_{k_{\tau}}  \star S_{h_{\tau\tau'}\acts k_{\tau'}}  \star \e_{h_{\tau\tau'}})(x_t^{\tau'}),
\eeq
The product is over the triangles of the 4-simplex, each of which shared by an (ordered) pair of tetrahedra.
$h_{\tau\tau'} \!=\!h^{\minus 1}_\tau h_{\tau'}$ is the parallel transport between $\tau$ and $\tau'$ and  $S_k(x):=\delta_{\minus k x^\minus k^\inv}(x^\plus)$  are the simplicity functions written in terms of the non-commutative delta functions (\ref{delta}). 

An important remark is in order, here. Without the simplicity functions, the bivectors $x^\tau_t, x^\tau_t$ would be identified up to (a sign and) parallel transport $h_{\tau\tau'}$ from  $\tau$ to $\tau'$, as they should be classically. 
However,  due to the specific properties of the star product, the presence of the simplicity functions  relaxes this identification. 
Using the plane waves expansion of $S_k$ and the definition of the star product on plane waves, one can indeed show that:
\beq
(\delta_{\minus x^{\tau}_t} \star S_{k_{\tau}}  \star S_{h_{\tau\tau'} \acts k_{\tau'}}  \star \e_{h_{\tau\tau'}})(x_t^{\tau'}) = 
\int \extd u^\tau_t \extd u^\tau_t ( \delta_{\minus x^{\tau}_t} \star S_{k_{\tau}}  \star S_{h_{\tau\tau'} \acts k_{\tau'}}   \star\e_{u_t^\tau h_{\tau\tau'} u_t^{\tau'}})(x_t^{\tau'}) 
\eeq
where the integration of the right hand side is over the product ${\SO(3)_\tau\times \SO(3)_{\tau'}}$ of stabilizer groups of the normals $k_{\tau}$ and $k_{\tau'}$. 
The fact that the amplitude does not force the complete identification up to parallel transport of the bivectors accross neighboring tetrahedra, also noticed 
in \cite{eteravalentin} and argued to be the main shortcoming of the Barrett-Crane model, appears, in our formalism, to be due to the star product structure, and thus the quantization map chosen for the constraints. 
As we argue also in the discussion of Sec. \ref{discussion},  such relaxation of parallel transport could be purely quantum effects, devoted to disappear in a suitable semi-classical limit. In fact such a requirement serves as a guide for the very definition of such a limit, which should reach a regime in which the star product is approximated by a commutative product. A scale $\lambda >0$ can be  introduced in the amplitude via a modification of the group Fourier transform, 
where $\SU(2)$ group elements are parametrized as $g\!=\!e^{\lambda \vec{p}\cdot \vec{\tau}}$ with $\vec{p} \!\in\! \R^3$ and $|\vec{p}|\!<\!1/\lambda$,
plane waves are replaced by $\e^\lambda_g(x) \!:=\!  \e^{\frac{i}{\lambda^2} \Tr xg}$, and the star product preserves the parameter \cite{PR3}: 
\[
\e^{\frac{i}{\lambda^2} \Tr xg} \star_{\lambda} \e^{\frac{i}{\lambda^2} \Tr xg'} = \e^{\frac{i}{\lambda^2} \Tr xgg'} 
\]
In the limit $\lambda \to 0$, corresponding to a regime of low curvature, the star product can be approximated by the standard pointwise product of functions on $\so(4)\!\sim\!\R^6$. In this parametrized framework, the parameter $\lambda$ appears not only in front of the action in the path integral, making it suitable to define the semi-classical regime, but also as a deformation parameter of the product used for the definition and the imposition of the constraints; 
this contributes to further quantum corrections, of which, the above relaxation of parallel transport may be a manifestation. 
This possibility will be studied in more detail elsewhere. 

Let us also point out that, if one defines, in the parametrized framework,  the $\su(2)$ non-commutative delta function  as 
$\delta\!=\!\frac{1}{\lambda^6} \int \extd g \e_g^\lambda$, then they formally become true delta function on $\su(2)\!\sim\!\R^3$ in the limit $\lambda \to 0$.  
With simplicity functions defined as before $S_k(x)\!=\!\delta_{\minus k x^\minus k^\inv}(x^\plus)$, this means that, in this limit, the constraints  in (\ref{vertex})
impose, for each triangle $t$, the following condition on the bivector $ x_t:=x_t^{\tau'}$ and the connection:  
\[
 - x^\plus_t = k_\tau x^\minus_t k_{\tau'} =  [h_{\tau\tau'}\acts k_{\tau'}]  x^\minus_t [h_{\tau\tau'}\acts k_{\tau'}]^{\minus 1}
\]
In words, the normals of the tetrahedra sharing $t$, expressed in the frame of $\tau$, differ from an element  in the stabilizer group of the $x^\minus_t$, namely
$ [h_{\tau\tau'} \!\acts\! k_{\tau'}] k^{ \minus 1} _\tau \in \U(1)_{x^\plus_t}$
This expresses the geometrical fact that the two normals belong the plane co-orthogonal to bivector $x_t$ (namely orthogonal to his Hodge dual, 
which defines the plane spanned by the triangle). 
In time gauge $k=1$, this becomes a condition on the boost part of the parallel transports $h_{\tau\tau'}$ between adjacent tetrahedra. 

\

The integration of the amplitude (\ref{vertex})  over half of the bivectors variables, namely, for each triangle $t \!=\!\langle\tau, \tau'\rangle$,  the variable $x_t^{\tau'}$ gives a function over the normals and the ten remaining  variables 
$x_t\!:=\!x_t^{\tau_t}$, which are bivectors of the triangle $t$ in its reference tetrahedron $\tau_t$. The resulting amplitude takes the form of a discretized (and non-commutative) path integral for constrained BF theory,  for a single 4-simplex, with fixed  boundary bivectors:
\beq
A(x_t; k_\tau) = \int \prod_t \left[\extd g_t e^{i \Tr x_t g_t } \star S_{k_{\tau_t}}(x_t)\right] \star
\int \prod_\tau \extd h_{\tau}  \prod_{t} S_{h_{\tau\tau'}\acts k_{\tau'}}(x_t) \, \delta(h_\tau   g_t  h_{\tau'}^{\minus 1})
\eeq
In this expression, we made explicit the part of the constraint  implementing the linear simplicity of the bivectors in their reference  tetrahedron,
and the  part (also dependent on the bivectors) entering the definition of the measure over the connections. 
Note that the closure constraint for the bivectors is only implemented after integration over the normals. 
That a proper spin foam model should have this form, with the corresponding constraints modifying the measure over the connections, has been advocated in \cite{sergei}.


\section{A model for the topological sector of Plebanski gravity}

In this section, we sketch the construction of a GFT model for the so-called {\sl topological} sector of Plebanski gravity as a constrained BF theory, where the constraints on the B-field impose that it can be expressed in terms of 1-forms  $B^{IJ} \!=\! e^I\wedge e^J$ rather than $\epsilon^{IJKL} e_I \wedge e_J$.
At the discrete level, their are implemented by requiring the same condition as in (\ref{linsimp}), but for the Hodge dual to B rather than B itself. 
In terms of B, the condition differs from (\ref{linsimp}) by a sign: there exists a $k\in \SU(2)$, such that $\forall j,  B_j^\plus = k B^\minus_j k^{\minus1}$. 

This condition  is obtained from the BF GFT in its extended formulation, by using an operator,  acting on the extended field (in metric variables)  as 
$\vphihat_k \mapsto (\hS^{top}\vphihat)_k:=\widehat{S}^{top}_k \star \vphihat_k$, 
where $\widehat{S}^{top}_k$ are the functions of  four $\so(4)$ variables, labelled by a $\SU(2)$ element $k$, defined in terms of the non-commutative $\delta$ functions (\ref{delta}) as 
\beq \label{Toposimp}
\hS^{top}_k(x_1, \cdots x_4) =  \prod_{j=1}^4 \delta_{kx^\minus_j k^\inv}(x^\plus_j)
\eeq 
Upon group Fourier transform, 
\[
\widehat{S}^{top}_k \star \vphihat_k= \int _{\SO(4)^4} [\extd g_j]^4 \, (S^{top}_k\vphi_k)(g_j) \,  \e_{g_1} \cdots \e_{g_4}
\]
this simplicity operator acts as:
\beq \label{Stopgroup}
(S^{top}_k \vphi_k)(g) := \prod_j \int_{\SO(3)^4_k} [\extd u_j]^4 \, \vphi_k(k^\inv u_j^{\minus 1} k g^\minus, u_j g^\plus).
\eeq
where we used the decomposition $g = (g^\minus, g^\plus)$ into self dual and anti-self dual components. 

Note that, unlike its analogue in the gravitational sector, the simplicity operator does {\sl not} define a projector\footnote{There are ways to impose such constraints via a projector, for e.g by acting on the field with $\hS^{top}_k$ 
by right star multiplication on the components $x^\minus$,  left on the components $x^\plus$; we will not investigate them in this paper.}. 
One can however check that it commutes with extended gauge invariance projector $\hC$  in (\ref{extgauge}). 
This is because, just like their analogues $\hS_k$, the simplicity functions $\hS^{top}_k$ satisfy the property that, given $h\in \SO(4)$, 
\beq \label{Topocommutation}
\e_h \star \widehat{S}^{top}_k = \widehat{S}^{top}_{h \rhd k} \star \e_h
\eeq
This property allows to impose the simplicity constraint {\sl covariantly}. 

A model can be defined by the action having the same form as (\ref{action}), but now defined in terms of fields $\hPsi =\hG^{top} \vphihat$, with $\hG^{top} \!=\! \hS^{top} \hC \!=\! \hC \hS^{top}$. 
Just as in the gravitational sector, the Feynman amplitudes take the form of simplicial path integrals for a constrained BF theory:
\beq \label{amptopmodel}
\cA_\cG = \int \prod_{<\sigma \tau>} \extd h_{\sigma \tau} \prod_t \extd^6x_t \left[\prod_t \vec{\bigstar}_{j=0}^{N_t}  \, 
\delta_{\bar{h}_{0j}x_t^{\minus} \bar{h}_{0j}^\inv}(x^\plus_t)\right] \star e^{i \sum_t \Tr \, x_t H_t}
\eeq
where the notations are the same as in the previous section: for each triangle $t$, the integers $j \! =\! 0\cdots N_t$ labels the tetrahedra sharing $t$, 
$h_{0j}$ is the holonomy between  the tetrahedra $0$ and $j$ and $\bar{h}\!:=\!h^\plus (h^\minus)^{\minus 1}$.
$\cA_\cG$ is the generic amplitude of closed graph expressed in time gauge, as (extended) gauge invariance allows to remove the dependency upon the normals.

The spin foam model can be calculated directly from its simplicial integral form, following the same route as in Sec. \ref{spinfoam}.
We will not describe in detail here the details of the model. 
Let us however have a look at the action of the simplicity operator  on the Peter-Weyl components on the fields. 
Gauge invariants  fields $\varphi_k$ expand into projected spinnetwork vertex functions  (\ref{spinbasis}): for $k=1$,
\beq \label{form}
\Psi^{(J_i, k_i, j)}_{m_i, m'_i}(g_i) =\left(\prod_{i=1}^4 D^{j_i^\plus}_{n_i m_i}(g_i^\plus) D^{j_i^\minus}_{n_i' m_i'}(g_i^\minus)
C^{j_i^\plus j^\minus_i k_i}_{n_i n_i' p_i}\right)  (\iota_j)^{k_i}_{p_i}
\eeq
We deduce from the orthogonality relations of the Wigner matrices
\[\int \extd u D^{j^\plus}_{mn}(u^{\minus 1}) D^{j^\minus}_{pq}(u) = \frac{1}{d_j^{\plus}} \delta_{j^\plus, j^\minus} \delta_{m, q} \delta_{n, p}\]
that simplicity (\ref{Stopgroup}) projects onto simple representations $j_i^\plus \!=\! j_i^\plus$, just like in the gravitational sector,  and exchanges 
the indices $n_i$ and $n'_i$ in $C^{j_i^\plus j^\minus_ik_i}_{n_i n_i' p_i}$ in the formula (\ref{form}).  
Given that the Clebsh Gordan coefficients satisfy:
\[
C^{j, j, k}_{m, n, p} = (-1)^{2j+k} C^{j, j, k}_{n,m,p},
\]
we conclude that $S^{top}$ projects onto simple $\SO(4)$  representation $J=(j,j)$, does not impose any restriction on the expansion of $(j,j)$ into $\SU(2)$ irreducible 
$k=0, \cdots 2j$, and acts on each component $(J, k)$ by multiplication by the phase $(-1)^{2j+k}$. 
 The expansion into the $\SU(2)$ representations $k$ is what defines the fusion coefficients charaterizing the new models \cite{newSF}, and in particular the EPR model. In the  EPR model, the expansion is restricted to spin $2k\!=\! j$. 
The model obtained here is manifestly different; the action of the simplicity function rather suggests a form analogue to the Ooguri model, 
as a sum over $\SO(4)$ representations of products of $(15j)$ symbols, where the representations are restricted to be simple.  

\

Thus, by following the same strategy as in the previous sections, a model for the topological sector of Plebanski gravity can be proposed in the non-commutative metric formulation of GFT, where the classical constraints are imposed via non-commutatice delta functions. Just like in the gravity case, the amplitudes are simplicial path integral with a clear geometric content. 
The resulting spin foam model is manifestly  distinct, at least in the full quantum regime,  from the one we would obtain by imposing the constraints on coherent states parameters \cite{newSF}.

\section{Discussion} \label{discussion}

In this section we summarize our results and re-examine in light of them the arguments against the Barrett-Crane model, by which, from now on, we mean the amplitudes associated to a given triangulation as obtained within the generalized GFT formalism presented above, whether written in the simplicial path integral or in the spin foam representation.

\ 

First of all, let us recapitulate the procedure we adopted, and the ingredients we put in. 

We started from a GFT formulation of quantum BF theory in 4 dimensions. The GFT field represents the quantization of the data associated to a single tetrahedron in a simplicial decomposition of spacetime (to be generated by the GFT perturbative expansion) in BF theory. Its arguments represent the corresponding simplicial BF variables. We extended the standard formalism to one including, for each tetrahedron, an additional variable valued in $S^3$, in preparation for the imposition of the geometric constraints that should turn this model into one for 4d gravity. We then imposed onto such extended GFT field an {\it extended covariance condition} affecting both the usual field arguments and the additional $S^3$ variable. This condition represents the gauge invariance constraint of (simplicial) BF theory, and introduces, when written in the form of an integral operator acting on a generic GFT field, the bulk gauge connection of the theory, encoding parallel transport among tetrahedral and simplex frames. In a geometric theory, this should also encode the {\it closure constraint} on the faces of the same tetrahedron. 

After appropriate constraints, in fact, the GFT field should turn into a quantum description of a (quantum) geometric tetrahedron.  The needed (simplicity) constraints can be split (both in the discrete and in the continuum setting) into two sets: diagonal and cross-diagonal simplicity, acting at the level of individual tetrahedra, and here imposed in their combined linear form \cite{newSF,danielesteffen}, and \lq volume\rq simplicity constraints. The variables on which these constraints  have to be imposed are exactly the Lie algebra elements appearing as arguments of the GFT fields in the metric representation, as confirmed by the GFT amplitudes written in the simplicial path integral representation. We then impose the first set of simplicity constraints, as well as the extended covariance (thus closure) as restrictions on the GFT field. Such restrictions then show up in the form of (non-commutative) delta function insertions in the simplicial BF path integral, at the level of GFT Feynman amplitudes, for all tetrahedra in the associated triangulation, and in the frames of the tetrahedra, thus result in a covariant imposition of the constraints. Moreover, we know \cite{newSF,danielesteffen} that the volume simplicity constraints follow automatically when parallel transport conditions among tetrahedral frames, simplicity and closure conditions hold. In turn, when all of these conditions hold, an assuming non-degeneracy, the set of bivectors define a unique geometry for the simplicial complex, and thus can be inverted for the set of edge lengths. No classical geometric condition is thus missing in the construction.

The next step, however, is to quantize these conditions. As mentioned, this process is necessarily ambiguous and a choice of quantization map must be made. And as mentioned, our procedure relies on the non-commutative Fourier transform and on  the resulting star product used to compose the building blocks of the simplicial path integral, including the simplicity constraints and the parallel transport conditions. Thus we have implicitly used, in the path integral setting, the Duflo quantization map. Once more, support for this choice, beside the more abstractly mathematical one, comes from the results of its application to pure BF theory \cite{aristidedaniele,GFTdiffeos} as well as to simpler systems \cite{Matti} (and, more recently, to canonical 3d gravity with cosmological constant \cite{alexdaniele}).

In this quantization, the closure condition of bivectors associated to the faces of each given tetrahedron is not manifest, in contrast to the standard formulation of BF theory, due to the extended form of gauge covariance. One may then want to include it explicitly, inserting a non-commutative delta functions at the level of the Feynman amplitudes of the model. However, as noticed also in \cite{thomasmuxin}, and remarked above, it is immediate to see that such insertion would be redundant and it could be re-absorbed in the integration over the discrete connection. The closure constraint is indeed already implemented by this integration.

As we have shown in the previous sections, this quantization procedure also implied a characteristic interrelation between simplicity constraints and parallel transport, which leads to the covariant imposition of the first, resulting restrictions on the connection, and a consequent relaxation of parallel transport conditions. This fact is not affected by the presence or value of the Immirzi parameter, but once more is a necessary consequence of the quantization chosen. Actually, it seems to go beyond it and be a feature of the other spin foam models as well. The result is that the configurations summed over in the simplicial path integral, and thus in the spin foam model, are not {\it strictu sensu} classical geometries, due to unavoidable quantum corrections (it can also be understood to be a consequence of the specific path integral measure that our GFT procedure defines). 

The end result, in this case, is a simplicial path integral with non-commutative variables, and in spin variables, a specific version of the Barrett-Crane spin foam amplitudes.

\medskip

Let us then reconsider the issues concerning the quantum simplicial geometry behind the BC model, including the one mentioned above, as made manifest in the metric/simplicial gravity path integral representation. These amounted to:

a) 4-simplices speak only through face representations, i.e. triangle areas \cite{asymBC};

b) bivectors associated to same triangle in different simplices are not identified (after parallel transport) \cite{eteravalentin, newSF};

c) normal vectors to the same tetahedron seen in different 4-simplices are uncorrelated \cite{eteradanieleCoupling,aristidedaniele};

d) the simplicity constraints are imposed in a non-covariant fashion \cite{aristidedaniele};

e) because of this non-covariance, there are missing constraints over the connection variables \cite{aristidedaniele}.

\medskip

The extended GFT formalism (with linear constraints and no Immirzi parameter) that we presented were meant exactly to solve the last three problems above, and they do. 

The other points refer to the coupling of simplices and to the coupling of tetrahedra within simplices. The non-commutative metric formulation shows that the bivector variables associated to the ten triangles in each 4-simplices are correctly identified across different 4-simplices sharing the same tetrahedron (and thus the same triangles), thus addressing the first issue. 
The additional coupling of normal vectors, in turn covariantly related to the bivectors by the simplicity constraints and by the generalised closure relation, ensures that the correlations among simplices resulting from bivector identifications are not undone by the lack of correlations of normals or by missing conditions on the discrete connection. 
Obviously, if one expands the transition amplitudes by Peter-Weyl into (simple) $\SO(4)$ representations and then integrates out {\it all variables except the representations themselves} (interpreted as quantum numbers for the areas of the triangles), the amplitudes associated to individual 4-simplices will share only such remaining variables with one another. This simply means that the whole simplicial geometry behind the model, and the exact correlations among geometric variables (bivectors and normals) in the same tetrahedron as seen by different (neighboring) 4-simplices have been encoded in a not-at-all-transparent way into the lower dimensional amplitudes.

\medskip

As we illustrated in Sec. \ref{sec:boundarystate}, 
point b), argued in  \cite{eteravalentin}, to be a serious  shortcoming of the Barrett-Crane model, has to do exactly with the relaxation of parallel transport resulting from the interplay between simplicity constraints and gauge covariance, due to the properties of the non-commutative star product. 
In fact, using the decomposition of the simplicity constraints into plane waves, the star product allows to recombine them with the plane wave corresponding to the unconstraint BF action, into a single effective action for the model. This is the action used also in \cite{eteravalentin}.  Now, as observed in this work, the  equations of motion for action do not force the complete identification up to parallel transport of the bivectors across neighboring tetrahedra. 
Is this a problem or a feature of the model, i.e  of the quantum amplitudes? We argue that it is a necessary feature -- which will be present also in any extension of the same construction involving the Immirzi parameter--, an indication of what quantum geometry may be, and not necessarily the sign of a problem with it. 
Our view is that the effective action obtained using the star product should be interpreted as an effective quantum-corrected action indicating properties of the quantum configurations summed over in the path integral, which indeed should not be expected to be classical simplicial geometries, in general. As a consequence the variations of the same action should not necessarily be interpreted as encoding the classical geometry behind the model, and their failure to reproduce classical simplicial geometry should not necessarily be interpreted as indicating that geometric conditions are missing among the ingredients leading to it. It just indicates how these necessary ingredients are affected by the quantization procedure.
While we leave a detailed analysis of the semi-classical expansion of the amplitudes for future work, we can already argue that such expansion will involve approximating the star product with a commutative product, and thus a commutative limit of the equations of motion resulting from the effective quantum action appearing in the path integral. This is exactly what happens in the simpler case considered in \cite{Matti}. In this approximation, the issue with parallel transport of bivectors disappears and classical simplicial geometry would then be recovered.

\medskip

In light of the above, the criticisms of (the first reference of) \cite{newSF}, saying that bivectors are not identified across simplices and that they are in the new models, therefore, can also be reconsidered. At a closer examination one notices that what is not identified across different simplices in the BC amplitudes are the {\it coherent state parameters} associated to each triangle in the  different 4-simplices, in quantum BF theory, on which a quantum version of the simplicity constraints is then imposed strongly (i.e. by means of delta functions) in the FK model (and in its extension to finite $\gamma$). The point is, however, that the coherent state parameters can be identified with the continuous bivector variables of BF theory, {\it only} in a semi-classical sense \cite{generalisedGFT}, i.e in the sense of mean values \cite{newSF} or in the asymptotic regime of the spin labels \cite{newSF, laurentflorian}. This is confirmed by explicit calculations of the group Fourier transform of Wigner representation matrices \cite{etera,fluxes,Matti}, in the coherent state basis, relating the coherent state parameters with the classical Lie algebra variables (for each $\SU(2)$ component).  This gives, for large spins $j$: $D^j_{\vec{n}\vec{n}}(\vec{x}) \approx \delta\left( \vec{x} - j \vec{n}\right)$.  It is directly the continuous bivector (Lie algebra) variables, on the other hand, that appear in our simplicial path integral representation of the amplitudes, and on which the simplicity constraints are imposed, by means of (non-commutative) delta functions, as appropriate in a path integral quantization \cite{thomasmuxin, generalisedGFT, aristidedaniele}. As for the identification of bivectors across tetrahedra within a given simplex, the same remark applies to the FK model (the other available one in absence of the Immirzi parameter): no strict identification up to parallel transport is imposed on the Lie algebra $B$ variables, and, as shown in \cite{eteravalentin}, even the identification of coherent state parameters is imposed only in a semiclassical limit.  

\medskip


\

We now move on to discuss the other arguments that have been put forward against the Barrett-Crane model. We do so for the sake of completeness, only. We have not dealt directly with them in this paper, nor worked in the context in which these concerns were raised. We limit ourselves, therefore, to reconsider the evidence suggesting these concerns, in light of what we learned from our results.

\medskip

As a preliminary step, we note a very general criticism raised \cite{thomasmuxin,generalisedGFT} against the BC model, which concerns other spin foam models as well. While the aim is to obtain a quantization of a classically constrained theory, i.e. the Plebanski formulation of gravity as a constrained BF theory, it results from a procedure that amounts to first quantizing the classical configurations of BF theory to get its quantum states ($\SO(4)$ spin networks) and then constraining such quantum states (usually in representation space). However, our metric representation of GFTs and spin foam models, providing their dual simplicial path integral representation, makes possible to follow a more standard path integral procedure and avoid some ambiguities, and to keep under control which of the constraints are imposed and how, even though it still (necessarily) depends on a choice of quantization map, as we discussed\footnote{The same motivation is shared by the work of \cite{thomasmuxin}, where however the non-commutativity of bivector variables, following from their conjugate nature to the non-abelian discrete connection and consistent with the canonical phase space of both BF theory and loop quantum gravity, is neglected.}.

\medskip

A popular criticism of the BC model is that it results from imposing the simplicity constraints \lq too strongly\rq \, at the quantum level \cite{newSF}. Indeed, when the constraints to be imposed are second class, as it is the case in the presence of Immirzi parameter or for quadratic simplicity constraints in general, they have to be imposed weakly in the quantum theory. On this, we notice two points. First of all, the model presented above imposes the simplicity constraints at the level of simplicial path integrals, by means of delta functions, as it is appropriate to do both for first and second class constraints. The nature of the constraints is manifested in their relation with other constraints and in the resulting further modification of the quantum measure. Second, while quadratic simplicity constraints are indeed second class, in the pure gravitational sector of Holst gravity or in the Plebanski theory {\it with linear constraints}, the constraints are actually first class \cite{robertojonathan}, so that they {\it have to} be imposed strongly. The only kind of \lq weakening\rq of constraints that is consistent with the classical theory in our case is the one following from the non-commutative nature of the bivector variables themselves (whence the dependence on a quantization map\footnote{One may also wonder why this quantization prescription is already present in the {\it classical} GFT action; this however is simply a confirmation of its interpretation as a \lq\lq third quantization\rq\rq formalism \cite{danielesteffen3rd}.}), which is taken into account by the $\star$-product and associated group Fourier transform.

\medskip

A related criticism \cite{sergei, sergei2} is that the BC model does not seem impose the secondary second class constraints that arise in the canonical analysis of the continuum theory \cite{canPleb,canHolst} (again when using quadratic simplicity constraints). This criticism applies as well as to all the new spin foam models \cite{newSF}. It is difficult to test any spin foam model in this respect, because none of them is derived following a canonical quantization procedure. Moreover, being based on a simplicial discretization, the standard canonical reasoning is difficult to apply and discrete counterparts of continuum secondary constraints are very difficult to define. A variant of this criticism adapted to the simplicial setting comes from the beautiful analysis of \cite{biancajimmy}, and from the standard discretization of the simplicity constraints of the classical theory, whether linear or quadratic \cite{BC,newSF,danielesteffen}. These analyses show that in the canonical setting it is necessary to impose {\it all} simplicity constraints, including so-called volume or edge simplicity constraints, on top of the ones we discussed explicitly, and also so-called \lq gluing constraints\rq which are analogue of secondary continuum constraints, in order to ensure geometricity of the classical configurations appearing in the model. This is indeed a reason for concern, and a detailed analysis is certainly needed in the context of the model we presented, and of the other available spin foam models. 
However, we emphasize that 
the non-commutative path integral formulation of spin foam models, to be developed also for the newer models, allows to study in a very direct way which constraints one is imposing in the definition of the spin foam model. Second, it is known \cite{newSF,danielesteffen} that, when imposed {\it at the covariant classical level in all tetrahedra} of the triangulation, the combination of diagonal, off-diagonal and closure constraints is enough to ensure full geometricity of the degrees of freedom included in the model, modulo degenerate configurations, if gauge covariance (parallel transport) is also imposed. 
Third, we have noticed how our covariant imposition of simplicity constraints involving the extra data encoded in the normals and the generalized form of the closure conditions imply further conditions on the normal vectors and on the discrete connection variables. The extra conditions affect the propagation of simplicity constraints from frame to frame across the triangulation, and thus their correct imposition in all tetrahedra, including those lying on different hypersurfaces. Thus it could be that such extra conditions are equivalent to the canonical secondary constraints. However, further work is needed to confirm or refute this possibility.

\medskip

Several criticisms concerned the boundary states of the BC model. Once more, more work is needed to define properly the boundary Hilbert space of the theory we presented. Still, our results offer some insight. It has been argued \cite{newSF,graviton} that states BC model do not have enough parameters to correspond to a 3-geometry. In other words, there would be a too-restricted set of commuting observables and in particular no label for the 3-volume (no intertwiner degree of freedom). In this respect we can notice that the boundary states of the BC model, as we have defined it, are given by graphs labelled by $\SO(4)$ representations (or Lie algebra elements (fluxes) \cite{fluxes} or group elements) on their links, subject to simplicity constraints, {\it and} one normal vector $k$ for each node, subject to our generalized gauge invariance condition. They are indeed projected spin networks as we have noticed \cite{sergei, sergei2, projected, eterasergei}. A heuristic interpretation of such state is as the discrete analogue of canonical wave functions on the extended gravity phase space, including lapse and shift vector: $\Psi(h, N, N^i)$. The standard states of the BC model, labelled only by simple representations and by the BC intertwiner, arise {\it after} gauge invariance is imposed and normals are integrated out. However, this integration over normals cannot be done freely (i.e. on each vertex separately, in the boundary spin network graph) because of the requirement of covariance, and because normals are correlated across tetrahedra by means of the discrete connection (also subject to constraints, as we have seen). It would rather seem that the states of the model, now including the normals, possess all the geometric information to be expected, while it can of course still be true that particular linear  combinations of states (as resulting for example from arbitrary integrations of normal degrees of freedom) do not correspond to well-defined boundary data.

\medskip

Clearly, this state space {\it does not} match the one of loop quantum gravity. This is based on $\SU(2)$ spin networks, that are labelled by an arbitrary $\SU(2)$ intertwiner, and that, while embeddable in a covariant way in $\SO(4)$ using indeed the projected spin network formalism \cite{sergei, eterasergei,maiteetera}, are based on the Ashtekar connection that requires for its definition the introduction of the Immirzi parameter \cite{LQG,newSF} . Indeed, the stated failure of the BC model to reproduce the tensorial structure of the (lattice) graviton propagator \cite{graviton}, at a closer examination, could be interpreted simply as a mismatch between the data encoded in LQG spin networks and used to define both the background boundary states and the same graviton observables, and the boundary states (and consequent dynamical data) of the BC model, that, as we have shown, should be taken to be the above projected spin networks. However, no such matching can be expected for a theory that is independent of  the Immirzi parameter from the start.

\medskip

Finally, the inclusion of the Immirzi parameter is the most interesting feature of the new spin foam models \cite{newSF}. A generalization of the non-commutative metric formalism presented here to include the Immirzi parameter is possible and will be presented elsewhere \cite{new4dGFTImmirzi}, together with the resulting new model. Still, two general points must be noted. First, one could expect any new model constructed in this way to reproduce (one of) the new spin foam models only in a semi-classical approximation, having in mind both our results on the topological sector of the theory and the construction of the new models based on group coherent states. Second, at present our main reason to want the Immirzi parameter in a quantum gravity theory is its necessity in the LQG context. It would be good to have further confirmations, either at the classical or quantum level, of its importance to capture the correct physics of a quantum spacetime.

\medskip

Last, we note that an important open issue is the role that degenerate geometries play in all spin foam models, including the one we have presented. Not only they affect drastically the imposition of simplicity constraints, and thus the geometric nature of the configurations included in them \cite{biancajimmy, constrainedBF}, but they may end up dominating the quantum dynamics on entropic grounds \cite{asymBC2}. It is also crucial to stress, however, as done in \cite{graviton} that this issue has to be studied in the context of the computation of physical quantities and that the insertion of appropriate observables or boundary states can change dramatically the respective weight of proper and degenerate quantum geometries.

\section*{Conclusions}
We have used the non-commutative metric formulation of group field theories and spin foam models to define a model of 4-dimensional quantum gravity as a constrained BF theory, without Immirzi parameter. This involved a generalization of the usual GFT framework to include both Lie algebra or group elements, associated to triangles in the triangulation, and normal vectors associated to tetrahedra of the same. The generalization led naturally to projected spin network states and the associated covariance under Lorentz transformations. The model is {\it uniquely} defined thanks to the projector nature of the generalized {\it geometricity} operator we introduced at the GFT level, encoding both simplicity constraints and (generalized) gauge covariance. The resulting model, for which we exhibited both a complete simplicial path integral expression and a spin foam representation, turns out to be a variant of the Barrett-Crane model, characterized by specific lower-dimensional amplitudes.
We also presented a similar construction for the topological sector of the Holst action, and discussed its relation with the EPR spin foam model.
In light of the above results, we have then re-examined the arguments against the Barrett-Crane model(s), concluding that it can still be considered a plausible quantization of 4d gravity as far as the encoding of simplicial geometry is concerned, and that further work is needed to either confirm or refute its validity.

\section*{Acknowledgements}
We thank Sergei Alexandrov, Valentin Bonzom, Bianca Dittrich, Roberto Pereira, James Ryan and Carlo Rovelli for discussions and comments.
DO gratefully acknowledges financial support from the Alexander Von Humboldt Stiftung, through a Sofja Kovalevskaja Prize. 
AB gratefully acknowledges financial support from `Triangle de la Physique' (Palaiseau-Orsay-Saclay), through a Postdoctoral research grant.

\end{document}